\documentclass{elsart}
\usepackage[T1]{fontenc}
\usepackage{graphicx}
\usepackage{amssymb}
\usepackage{bm}
\usepackage{subfigure}
\usepackage{graphpap}
\begin{document}
\begin{frontmatter}

\title{
Electronic properties of guanine-based nanowires
}
\author{
A.\ Calzolari,\corauthref{cor}}
\author{R.\ Di Felice,
and E.\ Molinari}
\address{
INFM-S$^3$ - National Research Center on
nanoStructures and Biosystems at Surfaces, and Dipartimento di Fisica
Universit\`a di Modena e Reggio Emilia, I-41100 Modena, Italy
}
\corauth[cor]{Corresponding author.
{\it Fax} +39 059 374794; {\it E-mail} calzolari.arrigo@unimore.it}
\begin{abstract}
We present a first-principle study of the electronic and conduction properties
of a few classes of nanowires constituted of guanine (G) molecules, self-assembled in
different geometries.
We first analyze the effect of the vertical $\pi$-$\pi$ interaction in model G-stack columns.
Then, we exploit the results obtained from those models to interpret the features of realistic 
stacked and hydrogen-bonded structures, namely the guanine quadruple helices and the planar ribbons.
With respect to natural DNA, the different structures as well as the inclusion of 
metal cations, drastically affect the bonding pattern among the bases, introducing novel features in the electronic properties of the systems.
These supramolecular G-aggregates, alternative to DNA, are expected to show intersting properties for molecular electronics applications.
\end{abstract}
\end{frontmatter}
%
\section{Introduction}
The idea that single molecules and/or molecular arrays might be anchored to
inorganic supports to be integrated into electronic circuits has inspired many
scientists in the last decade. An intense research effort is being developed to
identify the suitable molecular candidates, exploit them to fabricate devices,
and test the device efficiency.
This is the basic task of molecular electronics,
which is the solution of choice proposed for the continued
miniaturization of electronic circuits beyond the downfall of the
Moore's law predicted for silicon-based electronics.

By virtue of their recognition and self-assembling properties,~\cite{lehn}
DNA molecules appear as particularly suitable candidates as 
components for electronic devices at the nanoscale.
DNA's natural function is information storage and transmission,
through the pairing and stacking characteristics of its constituent bases.
The idea that it may also carry an electrical signal
would be extremely interesting in the bio-chemical
environment, but would also be appealing for the exploitation of DNA
in the nanotechnologies. 

In 1962, D.D. Eley and D.I. Spivey~\cite{eley}  suggested that $\pi-\pi$ interactions between stacked
base pairs in double-stranded DNA could provide a pathway for rapid, one-dimensional charge separation.
The issue of charge migration in DNA has recently become a hot topic in {\em solution chemistry} experiments.~\cite{barton}
Most of the available experimental results concern the migration of positive charge carriers ({\em e.g.} radical cations) within DNA molecules in solution.
A major target for oxidation is guanine (G),~\cite{giese,sugiyama} the
natural nucleic-acid base (Fig. 1a) with the lowest ionization potential.~\cite{heller}
The discussion of the results obtained using different methodologies and systems focused on the dependence of the
charge-transfer efficiency on the length of a $\pi$-pathway, serving as a bridge between remote donor and final acceptor sites.
In very general terms, two different regimes
can operate in charge conduction through DNA molecules:
as a  function of the base sequence,~\cite{megger,barton1} both  short-range (a few \AA) and  long-range (up to $\sim200$~\AA) charge migration have been observed.
In particular, in guanine-rich sequences of DNA, the oxidation/trapping process can happen over a long distance between the donor and the acceptor sites.

On the other hand, the construction of {\em solid-state} DNA-hybrid devices is a very recent achievement.
Both the combinatory principles intrinsic to nucleic acids and their chemistry can be exploited in building
precise, miniaturized and locally modulated patterns, where
the drawing of functional arrays is obtained through a series of programmed chemical reactions rather than by physical handling
of samples.
The realization of wired architectures via self-assembly opens the way to scale  the size of the devices down to the molecular level.~\cite{me}
In principle, the DNA-based materials may be used either as conductive wires or as templates for other materials.~\cite{braun}
While important successes have been achieved
in the last years in the immobilization of selected DNA molecules onto substrates and between nanoelectrodes~\cite{seminario}
confirming their excellent mechanical properties,
a clear understanding of DNA conductivity and its microscopic origin has not yet been achieved.
A recent series of experiments, which provided the direct measurements of the {\em dc} conductivity in DNA-hybrid structure,
reports contrasting results:~\cite{rosa,dekker}
depending on base sequence, molecule length, environmental conditions, substrates, and electrode materials, DNA exhibits 
insulating character,~\cite{braun,depablo,dekker1} semiconductor-like transport characteristics,~\cite{danny,lee}
ohmic behavior,~\cite{fink} and proximity-induced superconductivity.~\cite{kusamov}

Given these contradicting reports, no definite conclusions on the intrinsic conduction properties of DNA can be presently drawn from solid state experiments. 
As found for the charge transfer rates measured in solution chemistry experiments, the electrical properties of DNA-based devices
are strongly dependent on the sequence of the base pairs:
in particular an abundance of guanines seems to  enhance the conductivity of the double helix~\cite{danny,kawai2}.
However, even in the cases in which charge transport is observed, the current flow is very low, 
so that DNA might not to be reasonably considered a good conducting wire.

In this paper, we present a few nucleobase-aggregates, alternative to 
natural DNA,  expected to show interesting properties
for molecular electronics applications. 
Based on the experimental results stated above,
we focus on well-defined molecular structures by restricting the base sequence 
to the guanine  base alone.
Self-recognition and self-assembly processes of G-aggregates are controlled by a unique  sequence of functional groups, which act as donors and acceptors
of H-bonds (Fig. 1a).
Among the bases of DNA, guanine undergoes extensive self-assembly, mediated by H-bonding between G's, to give 
different aggregation states (dimers,  quadruple helices, ribbons) depending on the solvent 
concentration and on the presence (or absence) of specific cations.~\cite{gottarelli,anna} 
With respect to native DNA, the different structures drastically affect the bonding pattern 
among the bases, introducing novel features in the electronic properties of the system.

In order to understand the complicated electronic properties of these materials,
we first focus on the effects of the $\pi-\pi$ coupling
in selected model systems, constituted of vertical stack of guanine (poly(G) columns). 
Then, we consider two realistic systems: the stacked G-quartet helices ({\em G4-wires} or {\em quadruplexes})~\cite{anna,luisi}
and the planar H-bonding G-ribbons.~\cite{gottarelli1} The latter are the building blocks of guanosine fibers, 
successfully exploited in the construction of nanodevices.~\cite{rinaldi,meII,cingolani}

By means of ab-initio calculations, we study the electronic and conduction 
properties of these biomaterials from a microscopic point of view.
We focus on the effects of the $\pi-\pi$ coupling
and of the hydrogen bonding on the electrical properties of the self-assembled guanine nanowires.
In particular, the question that we address is whether the electronic properties of extended G-based structures
can be described in terms of extended orbitals and account for a band-like mechanism for charge transport. 
%
%
\section{Method}
In our first principle study of the self-assembled guanine nanowires,
the electronic structure is described according to the Density Functional Theory (DFT)~\cite{dreizler}.
DFT is lately gaining a large credit in the scientific community as a reliable and accurate method to describe large-scale biomolecular 
aggregates~\cite{friesner} including guanine-based stacks~\cite{prb,apl,jpc} and native DNA~\cite{depablo,parrinello1,adessi}.
We include the Generalized Gradient Approximations (GGAs) to the exchange-correlation (XC) functional, in order to improve
the description of highly directional interactions, such as H-bonding in molecular systems.~\cite{parrinello}

The electron-ion interactions are simulated via ab-initio pseudopotentials for each chemical species (H, C, N, O, K). 
The electron wavefunctions are expanded in a basis of plane waves.
We present two set of calculations with slightly different technical details:
poly(G) columns and ribbons are treated using  BLYP \cite{blyp} XC-functional, norm-conserving 
(Troullier-Martins~\cite{tm}) pseudopotentials in the in the Kleinman and Bylander factorized form~\cite{kb}, 
and an energy cut-off of 50 Ry for the expansion of  the single-particle wavefunctions.
In G4-wire calculations, instead, we exploit PW91~\cite{pw91} XC-functional, ultrasoft~\cite{vanderbilt} pseudopotentials, and 
25 Ry energy cut-off for the wavefunction expansion.
Initial test on gas-phase molecules and simple G-based aggregates confirmed that the latter approach 
does not modify substantially the description of the electronic structure of the systems,
while allowing the treatment of larger unit cells.

We simulate one-dimensional (1D) crystals by periodically replicating guanine building-block aggregates along
the axis of the nanowire, and including  a thick layer of vacuum in the other spatial directions to avoid spurious interactions between
adjacent replicas of the system.

For each selected G-assembly, we optimize the atomic geometry and the electronic structure. We perform quenched
ab-initio molecular dynamics simulations, according to a total-energy-and-force scheme for the search of the local minima.
All the atoms are allowed to relax, until the forces vanish within an accuracy of 0.05~eV/\AA. For each 
structure, we obtain both the geometry and the corresponding single-particle electron energies and wavefunctions.

Based on previous experimental and theoretical results~\cite{barton,ye} about similar nucleoside aggregates,
indicating that the sugar-phophate backbone is not directly involved in the electron transport processes,
we focus on  pure G aggregates that,
with respect to the experimental structures,
neglect the external backbone (G4-wire), the sugar and the alkyl chains (G-ribbons), and the surrounding medium.
The substituent groups attached to guanines are replaced with H atoms (H$_R$ in Fig. 1a) to saturate the dangling bonds.
%
%
\section{Poly(G) model columns}\label{polyg}
In order to clarify the complex intramolecular interactions in nucleobase-aggregates, we started by investigating 
the role of $\pi$-$\pi$ coupling in simple toy systems.~\cite{prb} We considered several infinite periodic stacks of guanines in a {\em wire-like} assembly,
with different inter-plane configurations. We stress that these poly(G) columns
do not describe real systems existing in nature, but they are useful tools to understand the effects of $\pi$ interactions
as a function of the geometrical parameters. The results obtained with these theoretical models constitute a first step
towards the interpretation of the electronic features of
some real systems, {\em e.g.} guanosine fibers  and G-quartet nanowires.

We started our simulation on isolated {\em stacked dimers}, pairs of G's lying in parallel planes whose distance
in the perpendicular direction is an output of our calculations. 
We analyzed several configurations, characterized by the relative azimuthal rotation angle of the two G's in the pair,
around the axis identified by the stacking direction, and/or by an in-plane translation.
The different geometries realize different $\pi-\pi$ overlaps.
After relaxing the structure of the G-dimers, we determined the equilibrium interplanar distance to be used for periodic column calculations.
For all the computed configurations the vertical distance between consecutive stacked planes 
is 3.37~\AA, which is the  experimental distance observed in
the B-DNA and in the G-quadruplexes. 
The detailed description of all selected configurations and their properties are reported elsewhere.~\cite{prb}
In the following, we present the results for two extreme cases with a very large and a very small $\pi-\pi$ superposition.

In the first configuration (Fig. 2, left inset) the two G's of the repeated dimer are perfectly
eclipsed; this realizes the maximum of the $\pi-\pi$ superposition.
In the second one (Fig. 2, right inset), the relative rotation angle is 36$^{\circ}$
around a suitable axis so to mimick the typical configuration
of two consecutive guanines along a double strand of B-DNA. In this case, the $\pi-\pi$ overlap is very small.

The analysis of the electron properties reveals interesting features.
We calculated the band structure along the stacking direction for both geometries reported in Figure 2.~\cite{nota2}
In both cases, we obtained a wide energy band-gap -- strongly underestimated in DFT -- between the valence band maximum and the conduction band minimum.

In the eclipsed configuration (maximum $\pi-\pi$ overlap), the bands deriving from the Highest Occupied Molecular Orbital (HOMO) 
and from the Lowest Unoccupied Molecular Orbital (LUMO) are dispersive.
The calculated effective masses for holes and electrons (m$_h$/m$_0$=1.04, m$_e$/m$_0$=1.41)
are similar to those of inorganic wide band-gap semiconductors, such as group-III nitrides, for which
a band-like conductivity is demonstrated (in presence of doping).
The $\pi-\pi$ interactions along the stacking direction induce extended channels for charge migration
through a band transport mechanism.
This interpretation is in agreement with the character of electronic states near the energy gap: 
In the columnar arrangement, through the interplanar G-G coupling, the $\pi$-like HOMO of guanine (Figure 1b)
induces the formation of Bloch-type delocalized orbitals that extend along the stacking directions (Figure 3).
Thus, we conclude that the electronic properties of the eclipsed G-wire may support band-like electron transport
through the base stack (provided a suitable doping mechanism might be achieved).

The picture drastically changes if we consider the B-DNA type configuration (small $\pi-\pi$ overlap).
The energy bands are dispersionless (right panel of Fig. 2) and the effective masses are
very high (m$_h$/m$_0$=5.25, m$_e$/m$_0\to\infty$).
\noindent
The low $\pi-\pi$ superposition does not induce the formation of extended Bloch-type orbitals along the
stacking direction, and the electronic states are localized at the individual molecules. 
This configuration does not support a band-like conductivity, and has electronic properties typical of a  molecular insulator.

In synthesis, the electronic properties of
vertical stacks of G molecules are very sensitive to $\pi-\pi$ coupling: a large $\pi-\pi$ superposition is a necessary condition
to sustain a Bloch-like conduction along the stacking direction.

Although these poly(G) structures do not directly simulate any real molecule, they provide some general notions that are helpful
to interpret experiments on G-rich DNA molecules.
For instance, our results suggest that relative geometry of neighboring bases in B-DNA does not support
energy dispersion and orbital delocalization (Bloch-type conduction) in a native poly(dG)-poly(dC) oligomer,~\cite{danny} due to the small $\pi$-$\pi$ overlap.
However, one cannot exclude that the interactions with the backbone and the external environment, as well as
the thermal fluctuations, may locally change the base-pair superposition for a finite time, 
thus favoring an increase of the $\pi-\pi$ overlap and the formation of extended
orbitals delocalized over a few bases, and eventually generating polarons.

As a final note to the above discussion, we can state that band-like conductivity (due to dispersive energy bands)
can only give a small contribution~\cite{ye} (if any) to the phenomenon of long-range charge migration in DNA. Other relevant concurring
mechanisms ({\em e.g.} tunneling, thermal hopping, polaron drift) must be taken into account to obtain a comprehensive interpretation.
%
%
\section{G4-wires}
The poor conductive properties of DNA 
(as suggested by the experiments~\cite{rosa} and by the results of previous section) 
have driven the research towards novel biomaterials that 
might provide electrical properties suitable for molecular electronics,
while maintaining the addressability and structured arrangement that are so attractive in the native DNA.
We identify a possible  candidate in
a homoguanine DNA-derivative: the  G4-wire (or quadruplex).
Due to the low ionization potential of guanine, this form might be suitable to mediate charge transport by hole conduction along the helix.
This low-dimensional structure, widely studied for its biological properties,~\cite{williamson} exhibits a surprising high stability in 
different chemical conditions, and may form nanoscale (100-1000 nm long) ordered wires.~\cite{afm}

In the presence of appropriate metal cations (especially K$^+$ and Na$^+$) solutions of homoguanylic strands, 
and of some lipophilic guanosine derivatives self-assemble in right-handed quadruple helices (Fig. 4a),
constituted of stacked  supramolecular  tetramers, known as {\em G-quartets} (G4).
Each G4 (Fig. 4b) consists of four coplanar hydrogen-bonded guanines, arranged in a square-like configuration. 
The metal cations, hosted in the central cavity resulting from the stacking of the G-quartets (Fig. 4c),
play a leading role in the assembly and stability of the aggregate.
They also  call for the investigation of the effects of the metal-molecule 
interactions on the electrical properties of the nanowires:
such interactions are currently explored as a viable route to enhance the conductivity of DNA derivatives.~\cite{lee} 

The starting atomic configuration for our simulations 
was extracted from the X-ray analysis of the short quadruplex d(TG$_4$T).~\cite{luisi} 
In the definition of the unitary cell, we exploited the periodicity modulo three of the G-quartet stack (Fig. 4c): we simulated periodically
repeated G4-wires by means of three stacked tetrads and three intercalated K$^+$ cations as shown in Figures 4a,c. 
Details about the geometry and the thermodinamic stability of G4-wires are reported elsewhere.~\cite{apl}

We analyze of the electronic structure in terms of the molecule-molecule and the metal-molecule interaction.
In the description of the guanine-guanine coupling we can also distinguish between the in-plane and the out-of-plane interactions.
Indeed, the in-plane structure of G-quartet (Fig. 4b) introduces novel features related to presence of the H-bonds~\cite{prb,apl},
while the out-of-plane behaviour is controlled by the $\pi$-$\pi$ coupling along the stacking direction.
The results of the simulation show that H-bonding is not capable of modifying the internal configuration of the single guanine; 
the polar nature of the H-bonds does not involve the sharing of electronic charge as the covalent bonds do. 
The molecular orbitals of the guanines do not laterally couple to form an {\em in-plane supramolecular orbitals} for the whole tetramer;
the same behavior is observed in the isolated G4, as illustrated by the HOMO state plotted in Fig. 4d. 
We conclude that the planar H-bonds do not significantly affect the band-like conductivity of the whole helix.

In the previous section, we showed that 
$\pi$-$\pi$ coupling may give rise to delocalized Bloch-type orbitals,
whose band dispersion depends on the relative rotation angle
between nucleobases in adjacent planes.
In the guanine quadruplexes,
the square symmetry of the tetrads increases the spatial $\pi$-$\pi$ overlap with
respect to a segment of G-rich B-DNA (30$^{\circ}$ instead of 36$^{\circ}$), and one might hope to find an enhanced band-like behavior.
Instead, we find that the out-of-plane $\pi$ superposition is not
sufficient to induce the formation of delocalized orbitals
along the axis of the quadruplex.
Consequently,  the energy bands along the stacking direction are flat,~\cite{jpc} reflecting the fact that
the orbitals remain localized at the component molecules. 

However, the band structure displays the presence of manifolds, due to 
the intermolecular interactions that split the energy levels.
The electron orbitals associated to a manifold have identical character
and localize on the constituent G molecules of the cell.
The energy levels in a multiplet are separated by an average energy difference of 20 meV;
therefore, even the coupling with a weak external perturbation ({\em e.g.} thermal fluctuations, electric field, etc.)
might be sufficient to allow the G-localized orbitals of a multiplet to interact,
producing an {\em effective delocalized orbital}.
This interpretation suggests a description of the electronic structure
where the spreading of the energy levels of the manifolds leads to the formation of {\em effective} band-like peaks
in the Density of States (DOS) (Fig. 5, black area).
The DOS of the G4-wire thus appears similar to those of wide-bandgap materials with  dispersive band-like structures.
We conclude that the coupling of the electronic states belonging to a manifold 
is coherent with the formation of {\em effective} extended electron channels, 
suitable to host mobile charge carriers along the wire.~\cite{jpc,wannier} 

The comparison with the empty G4 tube (without K$^+$ ions)~\cite{apl}
and with DNA structures available in the literature~\cite{depablo,parrinello} underlines how
the formation of {\em dense} manifolds (band-like peaks) is a general feature of the base-base interactions in
the H-bonded nucleobase stacks. 
Instead, the details of interplanar coupling depend on the specific arrangement (base sequence, rotation angle) 
and are influenced by the presence of metal cations intercalating the base stack.~\cite{jpc} 
Additionally, the effects due to the presence of  the cations depend on the specific choice of the metal species. 

In the case of potassium, the detailed description of its electronic structure (which includes the explicit treatment of 3p semicore shell) 
allows to investigate the metal-molecule interactions, neglected so far. 
We analyzed the metal-base coupling,
projecting the total density of states of the quadruplex (black area in Fig. 5) on guanine (red) and
potassium (green) atomic orbitals.
Whereas for a large part of the spectrum the whole contribution to the DOS stems only from the guanine states,
the inclusion of K$^{+}$ ions 
strongly affects the electronic properties in the energy region near the Fermi level (E=0 in Fig. 5),
as indicated by the superposition of G and K contributions in Figure 5.
Potassium contributes to the G4-wire electronic structure
with partially occupied  sp-orbitals, which derive from the mixing of the atomic 4s and 3p states.
The ionization process uniformly decreases of one unit the total number of electrons, without changing  
the character of these mixed states, which further couple with the HOMO's of constituent guanines giving rise to the {\em effective}-HOMO 
of the entire helix (marked with an arrow in Fig. 5).
The interaction between hybridized sp-orbitals of K$^{+}$ and the guanine orbitals accomplishes a double effect.
First, with respect to the empty G4 aggregate, the inclusion of potassium enhances the conduction properties of the 
system increasing the HOMO band width, and  generating further extended electron channels available for charge (hole) transport through the nanowire. 
Second, the rearrangement of the mixed K-G orbitals gives origin to a 
partially filled HOMO-derived peak, that characterizes the G4-wire as an intrinsic p-doped material.
The observation of delocalized holes at the topmost valence band is in agreement with the 
description of the electronic structure of potassium discussed above, where each  K$^+$ ion 
contributes to the whole electronic structure not with three fully occupied p-levels, but with four partially occupied sp-states.

We summarize the complex electronic properties the quadruplexes in Figure 6, 
where we report the contour plot of the HOMO-derived state of the wire
in two different planes.
The left panel of Fig. 6 has been obtained by slicing an isosurface of the HOMO charge with a plane perpendicular to the axis of the helix and
containing a G-quartet. The maxima of charge density (orange color) are localized around each guanine and no supramolecular orbitals are
visible across the H-bonds.
The in-plane charge distribution is basically the same of the isolated G-quartet plotted in Figure 4d.
The right panel of Figure 6 shows, instead, the contour plot of the HOMO-derived state in a plane parallel 
to the helix and containing the central metal cations.
We recognize a uniform distribution of the $\pi$ orbitals deriving from the single G's, but also the inner states stemming from the K$^+$ ions.
Two types of delocalized (Bloch-like) spatial electron channels can be identified. 
The first one, which is extended through the guanine core of the helix, is due to the molecule-molecule stacking interaction 
and corresponds to the formation of the effective band-peaks in the density of states. 
Similar channels are also observed for the other manifold-derived peaks.~\cite{jpc} 
Conversely, the second type of channel is specific of the HOMO peak, because it results from the metal-molecule interaction,
which is active only in the energy region near the top of valence band. 
This hybrid orbital is centered around the potassium ions in the central cavity of the wire, 
and it clearly shows the K$^+$ coupling with the coordinated oxygen atoms of the G molecules.

The above results, along with their attractive mechanical and self-assembling properties, suggest that the G4-wires 
should be explored as viable DNA-based conductors for nanoscale molecular electronics.
%
%
\section{G-ribbons}
As mentioned above, guanine derivatives may follow different self-assembly schemes, mediated by H-bonding. 
Besides the formation of the Hoogsteen dimers and four stranded helices,
a planar phase has been observed: in controlled experimental conditions, lipophilic guanosine derivatives self-assemble in a ribbon-like
structure, both in solution and in the solid-state fibers~\cite{gottarelli}.
The guanosine fibers~\cite{nota3} have been recently used to fabricate working nanodevices such as rectyfing diodes~\cite{meII} and transistors.~\cite{cingolani}

The planar ribbons are self-assembled quasi-1D structures of G's linked by a H-bonding network. 
We firstly studied the isolated planar ribbon,~\cite{physicaE} whose structure is presented in Figure 7.
To simulate extended ribbons we assembled coplanar G molecules according to experimental geometry~\cite{gottarelli1}.
In the same way as we did  for columnar G-stacks, we obtained the intermolecular distance by optimizing the structure of an isolated  dimer.
We used this value to fix the periodicity of the extended structures, whose geometry was also optimized.
The periodicity of the ribbons (1.1nm) is in good agreement with the experimental data (1.2nm).~\cite{gottarelli1}
The donor-acceptor distances (NH$\cdots$ N=2.9 \AA, NH$\cdots$ O=2.8 \AA) are consistent
with previous theoretical results for similar systems ({\sl e.g.} DNA base-pairs)~\cite{sponer}.
We find that the main effect of the H-bonding is to stabilize the structure, as proved by a large calculated energy gain
($\Delta E_{form}$=-820 meV/G).

The analysis of the electronic structure shows that, as it happens for the G-quartet, the H-bonds do not favor the formation of
{\em supramolecular orbitals} extended on the whole ribbon. On the contrary, electronic states remain localized around single G's.
Such electronic structure is not compatible with
a Bloch-like scheme for electron transport: in fact the energy bands along the axis of
the ribbon are totally flat, and the calculated density of state (not shown)~\cite{physicaE}
is characterized by sharp peaks corresponding to energy levels of the single guanines.
We conclude that an isolated planar ribbon should behave as an insulator,~\cite{prb,physicaE}
unless charge carriers migrate according to discrete hopping process.

A mechanism  concurring to charge transport may derive from the dipole-dipole interaction, which arises along the ribbon.~\cite{physicaE}
Indeed, G-ribbons have a net dipole oriented along their axes (Fig. 7);
it derives from the arrangement of the dipole moments of each guanine,~\cite{nota4}  which vectorially add up to
yield a finite and large spontaneous dipole moment lying parallel to the axis of the ribbon.
The associated electric field may favour the  charge motion through adjacent G's, by polarizing the medium.
It is worth noting that the presence of the macroscopic dipole is a consequence of the particular guanine arrangement
in the ribbon geometry. 
For instance, the square-like symmetry of the planar G-quartet (Fig. 4b) does not lead to a finite dipole for the whole tetrad.

Finally, to gain further insight into the fundamental interactions in extended 
gaunine aggregates, we studied  a 2D  crystal of stacked ribbons.
Following the results of Sec.~\ref{polyg}, we expect that the electronic properties of such systems vary as a function
of the relative base-base superposition.
Here, we present the results for a structural model in which the ribbons are perfectly eclipsed and separated by 3.4\AA along the stacking direction.
In principle, this geometry does not reproduce  the experimental solid-state arrangement of the ribbons:
Even though experimental results prove that the ribbons self-organize in fibers, 
the details of the stacking arrangement may depend on the length of the attached alkylic chains.
Thus, while our eclipsed model does not allow for a quantitative evaluation of the conduction properties of the real system,
it allows us to  underline  the relative role of the H-bonding and the $\pi-\pi$ interactions
when the $\pi-\pi$ coupling is maximized.

The electronic properties of the stacked configurations 
are simply the combination of two interactions: the H-bonding on the ribbon plane and the $\pi-\pi$ 
along the stacking direction. 
The band structure (middle panel of Fig. 8) shows two different behaviors
along the ribbon axis ($\Gamma-  X$ in the picture) and along the stacking direction ($\Gamma - A$).
In particular, as for the isolated ribbon, energy bands are almost flat along the axis ($\Gamma-X$);
instead, energy dispersion occurs along the stacking direction ($\Gamma-A$), as it happens for eclipsed poly(G) columns.
This is consistent with the analysis of single-particle orbitals shown in Figure 8.
Whereas  H-bonding does not favor lateral coupling, and the electronic states remain localized
at the G's (top panel of Fig. 8), the $\pi-\pi$ coupling induces the formation of
Bloch-type orbitals delocalized along the vertical direction (bottom panel of Fig. 8).

Stacked eclipsed ribbons present a wide-bandgap semiconducting behavior, compatible with a
band-like conduction in the stacking direction. 
As it happens with vertical columns, the band properties of stacked ribbons depend on the amount of the
$\pi-\pi$ overlap: different geometries with a lower $\pi-\pi$ superposition,~\cite{prb} are expected to have worse band-like
properties with larger effective masses with respect to the eclipsed configuration.
However, as a function of the superposition,
the $\pi-\pi$ interaction may enhance the band properties,
improving the mobility of charge carriers only along the stacking direction.

In summary, H-bonding and $\pi-\pi$ coupling play different roles:
H-bonding controls the in-plane self-assembly of guanines and stabilizes
the ordered structure, whereas $\pi-\pi$ coupling has a stronger influence on the out-of-plane electronic properties and determines
the contribution of delocalized band-like orbitals to the overall charge transport of G-rich systems.

Our results account for the 
experimental evidence of a semiconducting behavior of ribbon-like fibers,~\cite{meII}
where even a partial guanine overlap may favor the formation of band-like channels available for charge transport.
%
%
\section{Conclusions}
We presented a microscopic theoretical investigation of the electronic properties of a few selected nanowires
obtained by the different spatial arrangements of guanine molecules.

By an extensive study of guanine model solids,
we examined the characteristics of $\pi$-$\pi$ coupling and H-bonding, and discussed their effect
on the conductivity of the molecular assemblies.
We found that
Hydrogen-bonding does not support the band transport: 
no band dispersion is
present along  H-bonds in planar guanine aggregates, such as planar ribbons or isolated G-quartets.
Instead, base stacking is accompanied by
$\pi$-$\pi$ interactions that,
for sufficiently large overlap between adjacent $\pi$ orbitals, may induce energy dispersion and are
consistent with a finite contribution to coherent transport in the presence of doping or photoexcitation.
Therefore, band transport may be partially responsible
for charge mobility in nucleotide aggregates, in structures
characterized by a large base-base superposition: the electronic properties
of such materials are similar to those of inorganic wide-bandgap semiconductors.
These basic features occur in all the different  homo-guanine structures that we have
considered (poly(G) columns, G-ribbons, G4-wires).
However, the details of the electronic properties in each nanowire are strongly affected by the specific
supramolecular aggregation state and the coupling mechanisms
such as the metal-molecule hybridization in the quadruple helices  or the dipolar interaction in G-ribbons.

These features, along with the possibility of forming extended stacked wires at the
nanoscale length, make both the G4-wires and the G-ribbons appealing materials for the development of
biomolecular electronics, possibly more promising than DNA.
%
%
\begin{ack}
We friendly thank Anna Garbesi for her fundamental contributions and discussions.
This work was partially supported
by the EC through project "DNA-based nanowires" IST-2001-38951, by INFM through "Progetto calcolo parallelo"
which provided computer time at CINECA (Bologna, Italy), and by MIUR (Italy) through grant "FIRB-NOMADE".
\end{ack}
%
%

\newpage
%
%
%
%
\begin{figure}[!t]
\begin{center}
\includegraphics[clip,width=0.60\textwidth]{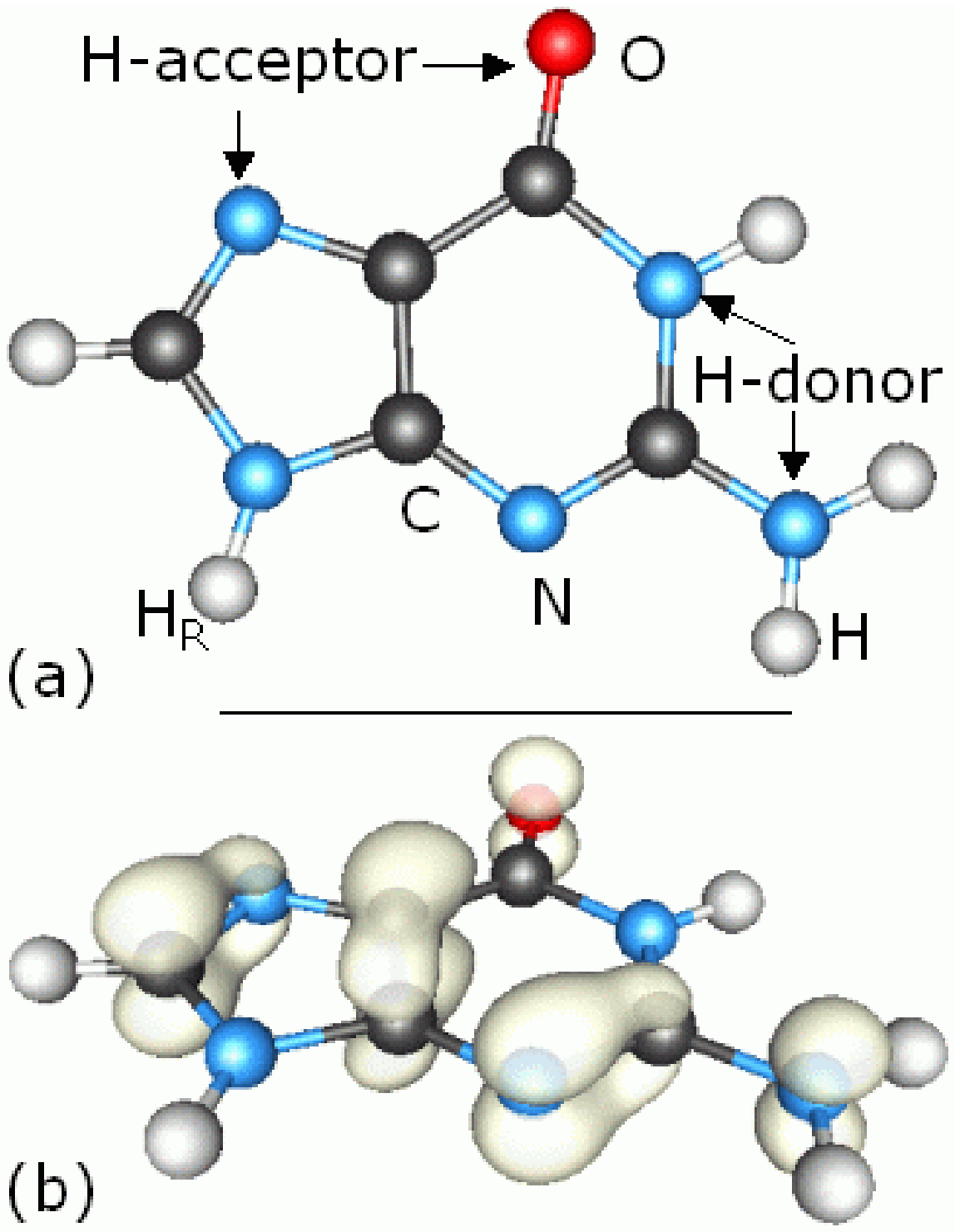}
\caption{
Isolated guanine molecule. 
(a) Atomic structure; (b) isosurface plot of HOMO state.
Different colors stand for different chemical species.
The hydrogen H$_R$ replaces the substituent groups attached to the molecule in the experimantal systems.
}
\end{center}
\label{gua}
\end{figure}
%
%
\begin{figure}[!t]
\begin{center}
\includegraphics[clip,width=0.8\textwidth]{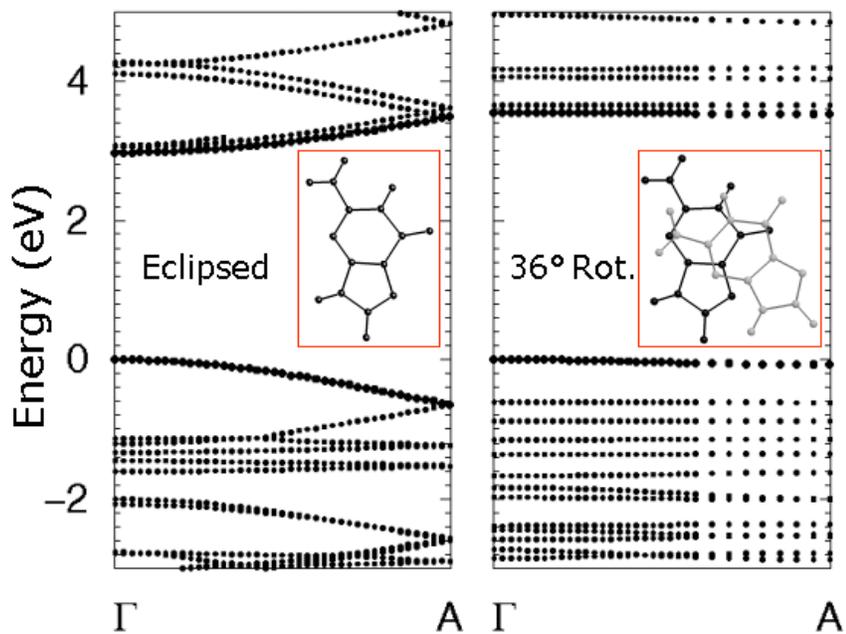}
\caption{
Band structure for eclipsed (left) and B-DNA type (right) geometries, calculated along the stacking direction.
The single particle energies reported in these plots are relative to the top of the highest valence band.
The insets show the top view of the two configurations respectively. 
}
\end{center}
\label{col_band}
\end{figure}
%
%
\begin{figure}[!b]
\begin{center}
\includegraphics[clip,width=0.60\textwidth]{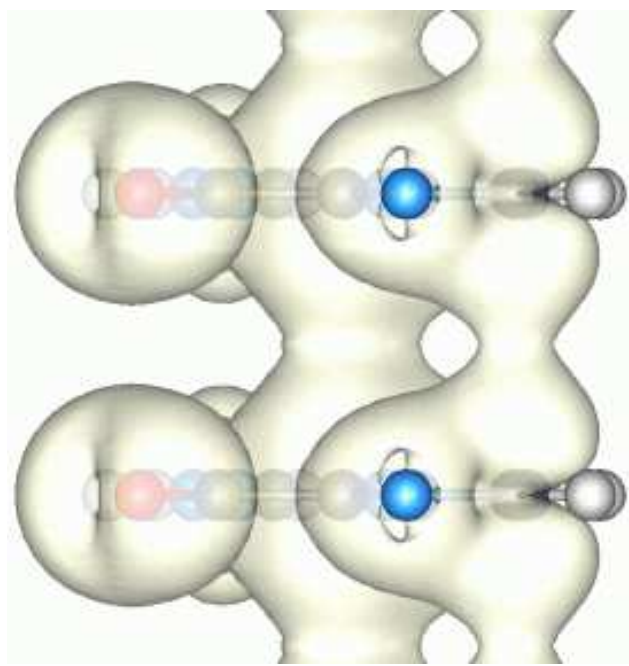}
\caption{
Isosurface plots of HOMO state for eclipsed poly(G) column, calculated at
the edge (A) of the BZ (side view).
}
\label{col_homo}
\end{center}
\end{figure}
%
%
\begin{figure}[!b]
\begin{center}
\includegraphics[clip,width=0.80\textwidth]{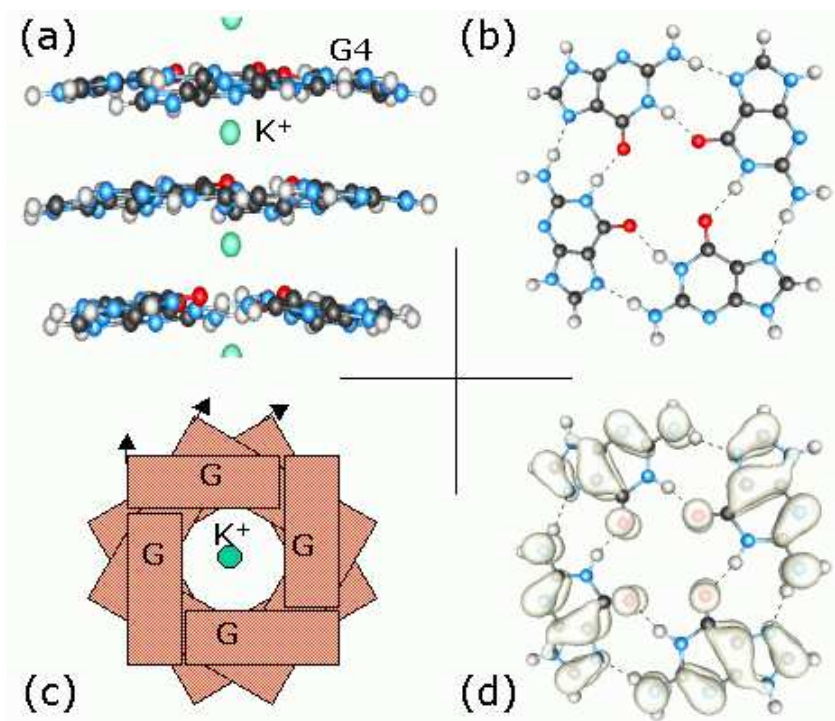}
\caption{
(a) Side view of periodically repeated G4-wire; 
(b) Geometry of isolated G-quartet;
(c) Schematic description of the periodicity modulo three in G-quartet stacks;
(d) Isosurface of the HOMO state of the isolated G-quartet.
Colors refers to different chemical species as in Fig. 1, green stands for K ions.
}
\label{3g4}
\end{center}
\end{figure}
%
%
\begin{figure}[!t]
\begin{center}
\includegraphics[clip,width=0.80\textwidth]{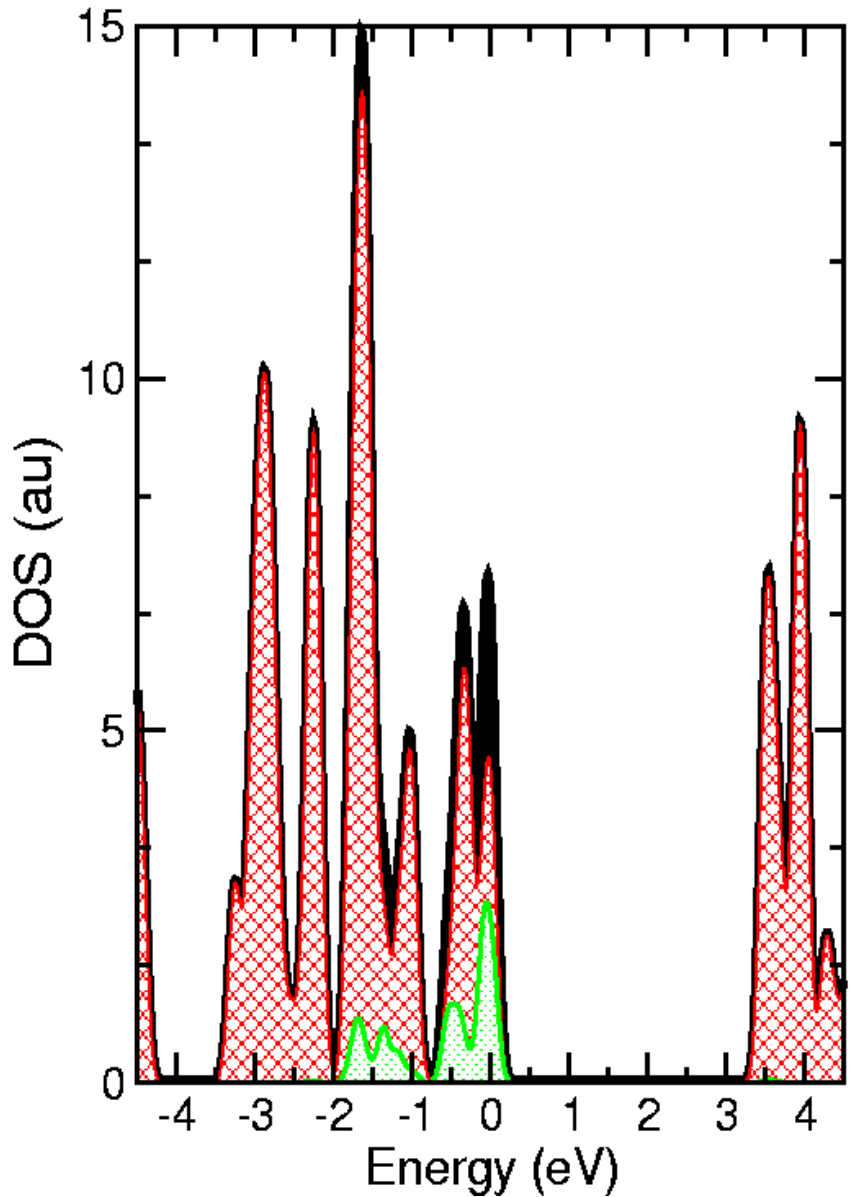}
\caption{
Total and Projected Density of States 
of the wire. The shaded black area represents the total DOS. The squared red (dotted green) area marks
the projection of the DOS onto G (K) contribution. Note that the total DOS and the G-DOS coincide in a large part
of the spectrum, where the red area almost coincides with the black one.
The zero of the reference energy scale indicates the Fermi level.
}
\label{3g4_dos}
\end{center}
\end{figure}
%
%
\begin{figure}[!t]
\begin{center}
\includegraphics[clip,width=0.80\textwidth]{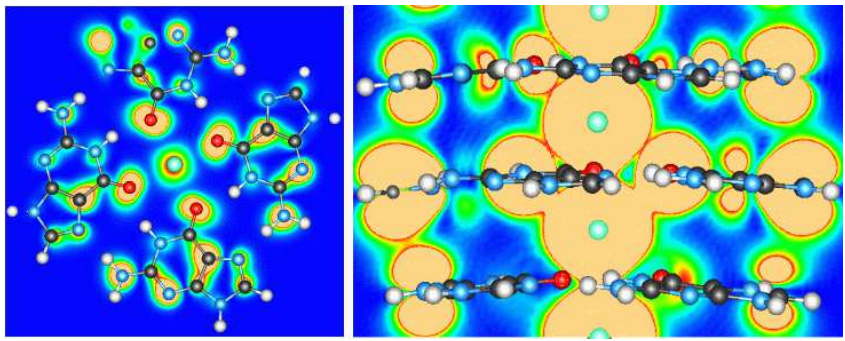}
\caption{Contour plot of the symmetrized charge density of the states belonging to the HOMO manifold, in a plane
perpendicular to the axis of the helix and
containing a G-quartet (left); and parallel to the helix and containing the central metal cations (right).
}
\label{3g4_stati}
\end{center}
\end{figure}
%
%
\begin{figure}[!t]
\begin{center}
\includegraphics[clip,width=0.80\textwidth]{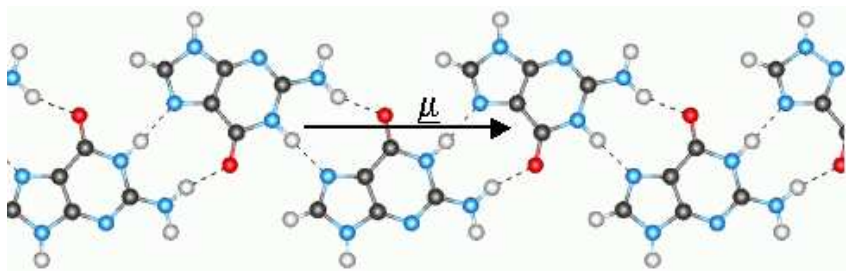}
\caption{
Isolated planar G-ribbon: relaxed geometry, and dipole moment ($\bar{\mu}$).
}
\label{nastro}
\end{center}
\end{figure}
%
%
\begin{figure}[!t]
\begin{center}
\includegraphics[clip,width=0.60\textwidth]{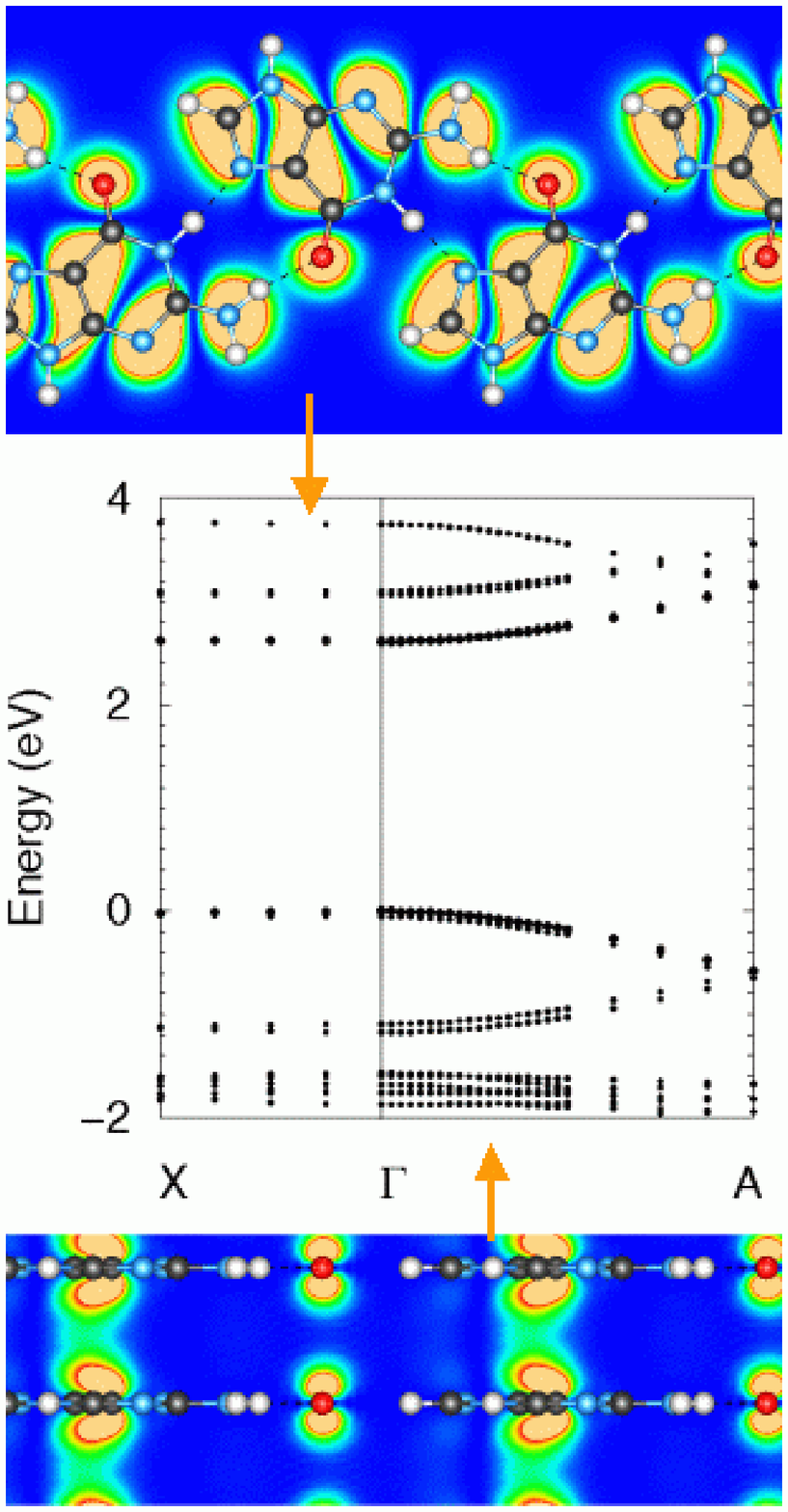}
\caption{
Central panel): Band structure of eclipsed stacked ribbons along the axis ($\Gamma-X$) and the stacking directions ($\Gamma-A$).
The zero energy is fixed to the top of valence band.
Isosurface of HOMO state in the plane containing the ribbon (top), and in a plane perpendicular to
the ribbons and parallel to the stacking direction (bottom).
}
\label{nastri}
\end{center}
\end{figure}

\begin{thebibliography}{00}
\bibitem{nota3}Guanosine fibers are obtained from lipophilic 3',5'-di-O-decanoyl-2'-deoxyguanosine in CHCl$_3$ solutions. 
               The deoxyguanosine molecule is constituted of the guanine base and the deoxyribose (DNA-sugar).
               In the experimental structure, lipophilic alkyl chains are attached to each deoxyguanosine in order to stabilize the structure.
               In our model, instead, we leave out the sugar and the attached alkyl chains.
\end{thebibliography}
\end{document}